\documentclass[11pt,reqno]{amsart}

\usepackage{amssymb}
\usepackage{latexsym}
\usepackage{amsmath}
\usepackage{amsthm}
\usepackage{amsxtra}

\usepackage{xcolor}
\usepackage{hyperref}

\usepackage{graphicx} 
\usepackage{float}
\usepackage{caption}
\usepackage{subcaption}

\title[Exploring Fermi's Paradox]{Exploring Fermi's Paradox using\\ an Intragalactic Colonization Model}

\author[G. Roudenko]{Gregory Roudenko}
\address{Cypress Bay High School, Weston, FL, 33331}
\email{roudenkog@gmail.com}

\author[Y. Pierre-Boyer]{Yurrian Pierre-Boyer}
\address{Department of Mathematics \& Statistics\\Florida International University,  Miami, FL, 33199}
\email{ypier009@@fiu.edu}



\date{}

\begin{document}

\begin{abstract}
        
We explore Fermi's Paradox via a system of differential equations and using simulations of dispersal and interactions between competing interplanetary civilizations. To quantify the resources and potentials of these worlds, three different state variables representing population, environment, and technology, are used. 
When encounters occur between two different civilizations, 
the deterministic Lanchester Battle Model is used to determine the outcome of the conflict. We use the Unity engine to simulate the possible outcomes of colonization by different types of civilizations to further investigate Fermi's question. 

When growth rates of population, technology and nature are out of balance, planetary civilizations can collapse. If the balance is adequate, then some civilizations can develop into dominating ones; nevertheless, they leave large spatial gaps in the distribution of their colonies. The unexpected result is that small civilizations can be left in existence by dominating civilizations in a galaxy due to those large gaps.
Our results provide some insights into the validity of various solutions to Fermi's Paradox. 

\end{abstract}

\maketitle
\tableofcontents

\section{Introduction}

\subsection{Fermi's Paradox}

Named after physicist Enrico Fermi, Fermi's Paradox questions why, given the high probability of advanced extraterrestrial civilizations in our Galaxy, we see no evidence of them. In 1950, Fermi and his colleagues at Los Alamos pondered this, with Fermi famously asking, ``Where is everybody?" \cite{Jones1985everybody}. Fermi calculated that we should have been visited many times over by extraterrestrial intelligences. There are two aspects of Fermi's paradox \cite{NatureAstronomy}:

(i) The first aspect of the paradox is the {\it grand scale} of our galaxy. There are approximately 200-400 billion stars in the Milky Way Galaxy and it is estimated that orbiting these stars there maybe as many as 300 million planets meeting conditions necessary for habitability of Earth-like organisms    \cite{Bryson_2021}, \cite{Forgot2013}, \cite{Tavares2020}.  Assuming that life occurs on even a small percentage of these planets, that's still a large number of planets that might carry life and perhaps a significant proportion of those that can spawn intelligent life may have developed on Earth-like planets with advanced civilizations in the past \cite{Howard_Smith2017}, \cite{Turner1985}.   

(ii) The second aspect of the paradox is the {\it probability of intelligent life} existing on other planets.  It is 
supposed that the high level of intelligence and the ability to manipulate the environment make it possible for them to overcome scarcity of resources and, at least locally, able to colonize new habitats. Those that are technologically advanced would be able to seek out new resources and colonize their star system or others. However, since evidence of extraterrestrial intelligent life has not yet been verified, the uncertainty remains and begs the question   \cite{Rospars2013}. Possible solutions are not too hard to imagine. For instance, other intelligent life forms may be rare or nonexistent, or our assumptions about the development or behavior about intelligent species are flawed, or our current understanding of the nature of the universe itself is quite incomplete \cite{Turner1985}. We further explore solutions to the paradox and how our model (in Section \ref{S:Unity}) can provide various outcomes, and possibly explain our own solo Earth civilization existence. 

Presupposing all that is necessary for the existence of technologically advanced civilizations, the question is what would happen, 
given a hypothetical Galaxy that has habitable planets with small but finite chances of spontaneous life under various scenarios?

\subsection{Fermi-esque Calculation}
What other considerations might have gone into Fermi's calculations? Let's suppose that the Milky Way Galaxy's diameter is about $10^5$ light years. At an average speed of $0.1$\emph{c}, where \emph{c} is the speed of light, it would take $10^6$ years to travel across it. Now, assume that the average distance between habitable planets is around four light years (a 40 year trip at $0.1$\emph{c} average speed) and it takes about 1000 years for an advanced civilization to colonize a primitive but habitable planet. Such a civilization on one side of the galaxy could traverse it not in a single leap but by small steps.  Dividing the diameter of the Galaxy by the length of the interplanetary distance and multiplying by the time it takes to colonize a planet and then adding the total time to travel across the Galaxy, it could conceivably take about $26$ million years to sequentially colonize and span it from one side to the other. 

The oldest known planetary system is Kepler-444 (five planets) aged about $11^{10}$ years \cite{Wall_2015}. If it took $5$ billion years to form an advanced civilization from the earliest beginnings of a planetary system $10$ billion years old, then the earliest advanced civilization may have appeared $5$ billion years ago. Dividing this five billion years between the spawning of the first advanced civilizations and now by the time it might take such an advanced civilization to colonize/transect the galaxy, we find that there is ample enough time for this crossing to have occurred 192 times even if one crossing did not begin until another ended. So, as Fermi questioned, {\it ``Where are they?" }  
\smallskip

\subsection{Drake Equation}

In 1961 the American astronomer Frank Drake proposed an equation to put together all the above pieces and quantify the possible number $N$ of radio-capable extraterrestrial life forms \cite{Burchell_2006}, \cite{Prantzos2013joint}. The equation reads:
\begin{equation}\label{Drake}
    N = R_* \, f_p \,  n_e \, f_l \, f_i \,  f_T \, L ,
\end{equation}
where \emph{$R_*$} is the rate of star formation in the Galaxy, \emph{$f_p$} is the fraction of stars with planetary systems, \emph{$n_e$} is the average number of planets around each star per solar system (that can support life), \emph{$f_l$} is the fraction of planets where organic life developed, \emph{$f_i$} is the fraction of planets where intelligent life developed, \emph{$f_T$} is the fraction of planets with technological civilizations, and \emph{L} is the average duration of the radio-communication phase from those civilizations \cite{Prantzos2013joint}. The last four terms are unknown, so only hypothetical estimates are possible. 

Although this equation has been used by optimists and pessimists (with greatly varied results), most arguments for using this equation suffer from overconfident reasoning when choosing specific numbers for likelihoods of events we have little knowledge about. For instance, the chance of bio-genesis on Earth-like planets is one where the probability ranges from near zero (we are alone) to near $1$ (bio-genesis is an inevitable and spontaneous process). Depending on the numbers chosen, one could speculate that there are thousands of civilizations in a Galaxy or that there are none.

\subsection{Hypotheses}\label{S:Zoo}

There exist various hypothesized solutions to the Fermi paradox to explain why exactly alien life has not been encountered yet. 
We recall 2 hypotheses that are well-known and we expect to occur in our simulations: 

\begin{enumerate}
    \item The {\it Great Filter hypothesis} \cite{greatfilter} posits that there is some extremely difficult stage that life needs to go through before gaining the ability to colonize space. This could be the formation of multicellular life, the evolution of an intelligent species, or perhaps something humanity has not experienced yet.
    \item The {\it Zoo Hypothesis} \cite{wikizoo} suggests that intelligent civilizations are widespread throughout the galaxy, however, have no interest in contacting or colonizing others and instead simply observe them from a distance.
\end{enumerate}

\subsection{Modeling and Results}
In this paper we explore Fermi's Paradox using a system of differential equations together with a model for disputes and later incorporate all of that into a Unity simulation engine, producing simulations of dispersal and interactions in multiple competitive planetary civilizations. To quantify the resources and potentials of these worlds, three different state variables representing population $P$, environment $E$, and technology $T$, are used. 
When encounters occur between different civilizations to colonize a given planet, the deterministic Lanchester Battle Model is used to determine the outcome of the conflict.  
We use the Unity engine \cite{Unity} to simulate the possible outcomes of colonization by different types of civilizations in order to try to give some answers to  Fermi's question. 

We find that when growth rates of population, technology and nature are out of balance, planetary civilizations can collapse. 
If the balance is restored or maintained, then some civilizations can develop into the dominating ones; nevertheless, we observe that even colossal civilizations leave large spatial gaps in the distribution of their colonies. The unexpected result is that {\it small civilizations can be left in existence} by dominating civilizations in a galaxy due to those large gaps. 
In our simulations, this was due to the large civilization entering a fixed state, where each of its populated planets ended up either with no colonization attempts left, no planets near them to
colonize, or with planets near them that simply had unsuitable conditions and kept dying out. Thus, our results provide some insights into the validity of various possible solutions to Fermi's Paradox.
\smallskip

This paper is organized as follows: in Section \ref{S:PET} we describe the PET model for a civilization on one planet, more precisely, the ODE model that describes time dependence of population, environment and technology values of the civilization occupying one planet. In subsection \ref{SS:Coloniz} we start putting planets together in a galaxy and explain a possible occupation or colonization of planets in that galaxy. In Section \ref{S:LBM} we discuss a basic dispute model, Lanchester Battle Model, which we use to describe the situation, where one civilization tries to occupy a planet with an existing other civilization. Finally, in Section \ref{S:Unity} we put together our models and simulations in {\it one unified model}, that will provide possible outcomes for Fermi's question. The simulation results are given in Section \ref{S:results} and we finish with conclusions in \S \ref{S:Concl}. 

\medskip

{\bf Acknowledgments.}
The research of this project initiated and was mostly done during the Summer 2023 REU program ``AMRPU @ FIU", which took place at the Department of Mathematics and Statistics, Florida International University, and was supported by the NSF (REU Site) grant DMS-2050971.  We would like to thank our faculty advisor Dr. Tennenbaum for suggesting the topic and helpful guidance. 

\section{Model Description}

\subsection{PET Model}\label{S:PET}

To simulate changes in population, advancement in technology, and the consumption of the surrounding environment by a civilization (on one planet), we use what is called a PET model. This model, consisting of three differential equations, is intended to simulate basic aspects of a civilization on one planet. Later we incorporate this model into the simulation of several plants in a galaxy, but for now we start with the PET model. 

\subsubsection{What is the PET Model?}

The PET model is based on 3 variables: \emph{P} - Population, \emph{E} - Environment, and \emph{T} - Technology.
The system of equations describing the model is given by

\begin{align}
\frac{dP}{dt} & = rP\Big( 1-\frac{P}{K} \Big) - vTP \label{population} \\
\frac{dE}{dt} & = \lambda - \bigg( aE + bP + c\, \Big(\frac{T}{g + T} \Big)P \bigg) \label{environment} \\
\frac{dT}{dt} & = \frac{m\,TEP^2}{h + TEP} - uT, \label{technology}
\end{align}
where \emph{P}, \emph{E}, and \emph{T} are the values of Population, Environment, and Technology at a given time $t$, respectively. The parameter $K$ is the carrying capacity, which is proportional to the product of the environment and technology variables, i.e.,  $K = f \, E T$, where $f$ is a constant; the rest of the lowercase variables being constants as well. 

In equation \eqref{population}, the first term describes logistic growth for the population $P$ with the carrying capacity $K$ as described above. The second term $v \, TP$ is a generalized model of the amount of lives lost due to infighting and other downsides of industrial technology such as pollution, drugs, accidents, and so forth.

In equation \eqref{environment}, the constant $\lambda$ is the natural speed at which the environment regenerates. The other factors inside of the parentheses represent everything that negatively impacts the environment, the first term $a E$ being natural and the last two terms being human-caused:
\begin{itemize}
\item
the first term $a E$ is the natural turnover rate of the environment, where $1/a$ would be the lifespan of one average organism used for consumption by the civilization;
\item
the term $b P$ is the consumption rate of the environment by the population, and it is linearly proportional to the amount of people consuming the environment;
\item
the last term, $c \, \frac{T}{g + T} \,P$, is the consumption rate of the environment by technology, rapidly growing initially but then asymptotically approaching a value proportional to the population $P$. 
\end{itemize}

In equation \eqref{technology}, there are two terms: one for growth of technology and one meant to symbolize ``infrastructure destruction" due to various things such as conflicts and needing to replace old infrastructure to the one with newer technology. The growth term for technology is, for small initial values, proportional to the square of population. This is because we view technology as a collaborative process where it takes the interaction of people to innovate because they need to share ideas. The denominator in this term limits the rate at which new technologies can form, limiting it to a value approximately proportional to that of the population. This prevents technology from blowing up and rapidly consuming all available resources.

\subsubsection{Simulation}
For a sample planet, to follow the time evolution of functions $P, E$ and $T$ via the PET model, we discretized the equations \eqref{population}, \eqref{environment}, and \eqref{technology} as difference equations via Euler's method and ran them in Python. 

We start at time $t=0$ and proceed with a time step $\Delta t = 0.001$ until the time $T=50$. 
The initial conditions for our functions are 
\begin{equation}\label{E:ID}
P(0)=1, \quad T(0)=1, \quad E(0) = 200. 
\end{equation}

In the equation \eqref{population} we use the following parameters: $r = 0.3$ (logistic growth constant of proportionality) and $v = 0.01$ (how much damage infighting does).
For the carrying capacity $K = f\, ET$, we take $f = 0.3$ as the carrying capacity coefficient.

In the equation \eqref{environment} we use the following 
coefficients: 
$\lambda = 20$ (biomass natural regeneration rate), $a = 0.1$ (biomass natural death rate: $1/a$ is how long it takes for environment to replace itself),
$b = 0.5$ (how fast population consumes environment),
$c = 1$ (how fast technology consumes environment),
$g = 100$ (parameter that tempers negative effects of technology).  

In the equation \eqref{technology} the parameters used are as follows: 
$m = 0.025$ (technology growth coefficient), 
$h = 500 $  (how quickly technology growth eventually slows down), 
$u = 0.01$  (how fast technology development is lost due to infighting).

\smallskip

\begin{figure}[htb!]
    \centering
    \includegraphics[height = 0.3\textheight,width=0.45\textheight]{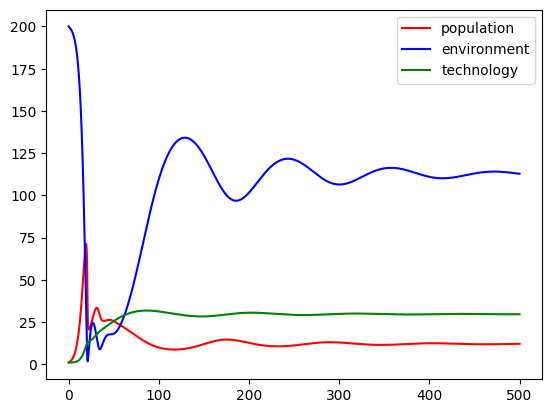}
    \caption{Progression of population $P$ (red), technology $T$ (green), and environmental conditions $E$ (blue) over time on one planet by PET model.}
    \label{F:PETprogress}
\end{figure}

\begin{figure}[htb]
    \centering
    \includegraphics[height = 0.3\textheight,width=0.45\textheight]{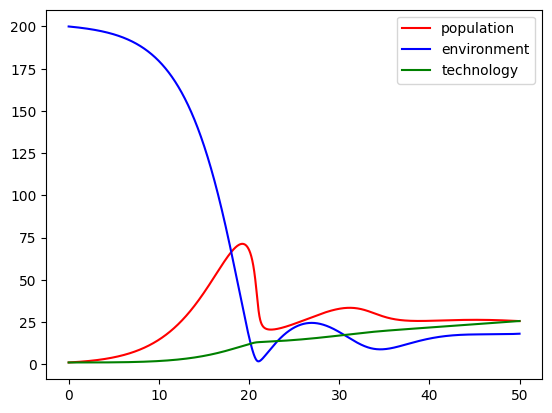}
    \caption{Zoom-in of the first 50 time-steps from Fig. \ref{F:PETprogress}.}
    \label{F:zoomprogress}
\end{figure}

The results of this simulation with the initial data \eqref{E:ID} is given in Figures \ref{F:PETprogress} and \ref{F:zoomprogress}. 
The population $P$ (red line) initially shows exponential growth as technology rapidly progresses, see Figure \ref{F:PETprogress}, but then levels off after about 100 time steps. 
The environment $E$ (blue curve) drops significantly down right away, however, later picks up and with oscillations starts leveling off. (This can be understood that initially with the population growth the environment is damages significantly, however, as the technology develops, hence the awareness of environment preservation, efforts are put into its restoration). The technology $T$ (green curve) grows initially and then levels off. One can observe that after about 150 time steps, all 3 values begin to stabilize.

To better understand the initial evolution that is occurring on the planet in the simulation shown in Figure \ref{F:PETprogress}, we show a zoomed in version of the first 50 steps in Figure \ref{F:zoomprogress}. 

Analyzing the Figure \ref{F:zoomprogress}, one could note that 
the current condition of the Earth would be somewhere between the 10\emph{th} and 20\emph{th} time-step, as population and technology rapidly grow and the environmental condition rapidly deteriorates. 

{\it Remark:} the $y$-axis scales in the simulation are arbitrary and are only meant to show the relative scale of each variable as time progresses. For example, the population being ``75" near time step 20 merely reflects that there are 75 times the number of intelligent organisms on the planet as there were initially.

\subsection{Colonization Simulation}\label{SS:Coloniz}

\begin{figure}[htb]
    \centering
    \includegraphics[width=.6\linewidth,height=.3\linewidth]{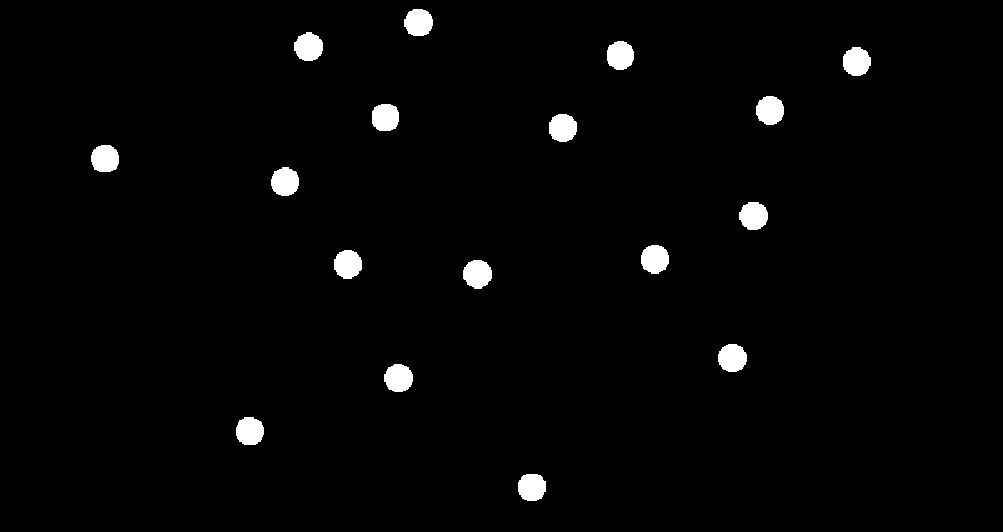}
    \caption{Randomly dispersed planets with no life yet (initial stage).}
    \label{nolifeplanets}
\end{figure}

Our next step is to consider planets together in some space, we call a galaxy.
We built the colonization simulator in the Unity real-time 3D development engine  \cite{Unity} as it allows to depict the simulations with visual effects and is one of the standard engines to do that. Here, each planet is depicted as a small circle (we often refer to it as a point or dot), randomly dispersed on a flat plane (to resemble a galactic plane) such that all points are at least a minimum distance \textit{d} away from each other and such that all points fall within a distance \textit{D} of a central point of our simulated galaxy (in other words, $D$ is the radius of the galaxy), see Figure \ref{nolifeplanets} for the initial set-up. A dot is white when the planet it represents is unoccupied, i.e., life has not spawned on it (or it is un-colonized).  

\smallskip

\begin{figure}[htb]
\centering
\includegraphics[width=.6\linewidth,height=.3\linewidth]{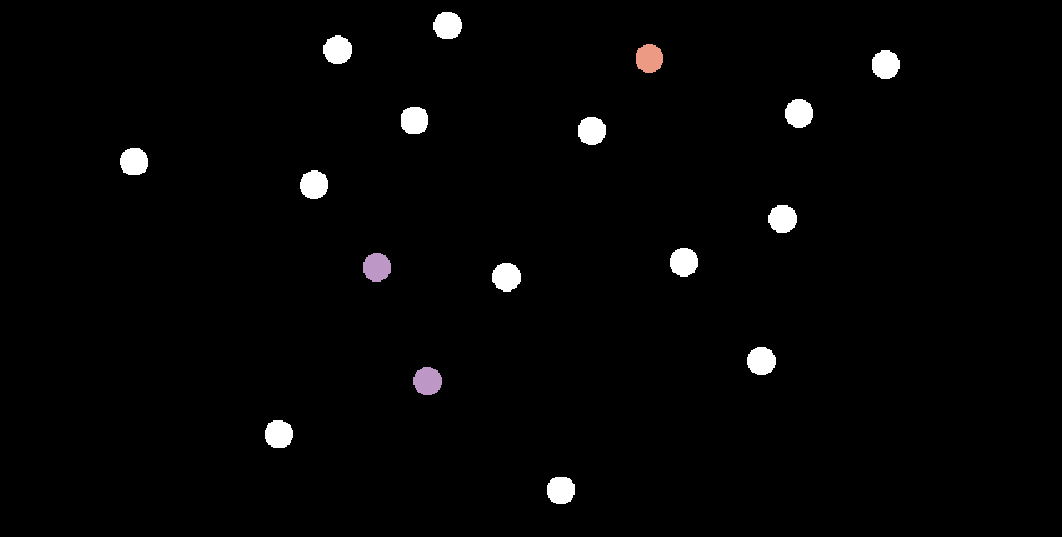}
\caption{The purple civilization has colonized a second planet, while the orange civilization still only consists of one planet.}
\label{colonizedplanets}
\end{figure}

We mark different civilizations with  different colors, for instance, orange and purple civilizations in Figure \ref{colonizedplanets}.
We proceed our simulation as follows: planets will colonize the nearest neighboring planet when they reach a minimum threshold of population and technology. In our simulations we use the following thresholds: technology is above 20, populations is above 15, and the time since the last colony sent is 5 or higher (this is the amount of time it takes to prepare an expedition). 

If a planet is unoccupied and a certain civilization reaches it within its threshold, the civilization will colonize the planet with the population of colonists, e.g., in Figure \ref{colonizedplanets} the purple civilization has colonized a planet in addition to its original home planet, thus, it has two planets. When a planet is colonized, we immediately start running the PET model to simulate new population growth as well as technology development and environmental changes on that planet. 
If the thresholds are not reached yet, only PET model is running for a specific planet, e.g., the orange dot in Figure \ref{colonizedplanets} represents the population that just occupied one planet.

We next investigate what happens if a civilization attempts to colonize a planet controlled by another civilization. For that we consider the following model for disputes. 
\smallskip

\section{Lanchester Battle Model}\label{S:LBM}

To simulate disputes between a civilization occupying a planet and another civilization attempting to colonize it, we implemented a simple deterministic battle model based on Lanchester's Law of Battle Strength, see \cite{Mackay2006Lanchester}, \cite{Lanchester1956}.
This model is defined by a system of linear equations, 
\begin{equation}\label{E:LBM}
\left\{
\begin{aligned}
\frac{dR}{dt} & = -g \, G\\ 
\frac{dG}{dt} &= -r \,R, 
\end{aligned}
\right.
\end{equation}
where \emph{G} and \emph{R} are the populations of the opposing armies, say, red and blue, 
and \emph{g} and \emph{r} are the coefficients of ``fighting effectiveness" or how many enemies one soldier can kill per unit \emph{t} of time. The values of \emph{g} and \emph{r} are based on the respective technology levels of the civilizations at war.
\smallskip

We call this model ``simple" because it makes strong simple assumptions: the model simplifies all interactions as battles, however, it is meant to take into account all forms of disputes between two civilizations. It also assumes two groups will quickly be at odds with each other and will ``attack" each other at a rate proportional to their numbers and a strength determined by their technical sophistication, \cite{Mackay2006Lanchester}, \cite{Lanchester1956}. More accurate warfare simulation models can be complex and stochastic, so this model does not completely accurately represent real warfare, however it provides a baseline for determining who has the advantage.
Another benefit of using such a simple model is that we can easily solve for the exact outcome of any dispute using the solutions for the coupled differential equations, that is,

\begin{equation}\label{E:LBMsol}
\left\{
\begin{aligned}
G(t) & = \frac{\sqrt{g} \, G(0) \, \cosh(\sqrt{rg} \, t)-\sqrt{r} \, R(0) \, \sinh(\sqrt{rg}\, t)}{\sqrt{g}}\\
R(t)  & = \frac{\sqrt{r}\, R(0) \, \cosh(\sqrt{rg}\, t)-\sqrt{g}\, G(0)\, \sinh(\sqrt{rg}\, t)}{\sqrt{r}},
\end{aligned}
\right.
\end{equation}
where $R(0)$ is the initial population of the red army $R$ and $gG(0)$ is that of the green army $G$.

\begin{figure}[htb]
\includegraphics[width=.5\linewidth,height=.3\linewidth]{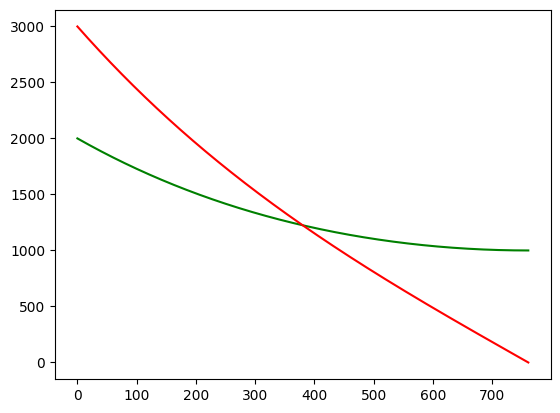}
\caption{\small Red colony has a larger initial size but smaller fighting effectiveness, while Green colony is smaller initially but with higher coefficient of fighting effectiveness. In this simulation, the Green wins (the horizontal axis is time), and hence, will colonize a given planet.}
\label{F:LCB-model}
\end{figure}
For a visual depiction of this model, we provide an example of two colonies, Red and Green, with the initial size of the Red colony $R(0)=3000$ and fighting effectiveness $r=0.1$, and the initial size of the Green colony $G(0)=2000$ and fighting effectiveness $g=0.3$. One can see in Figure \ref{F:LCB-model} that in this simple model, given the above initial data, the green colony wins (and thus, will colonize a given planet). 

\section{The Unified Model}\label{S:Unity} 

For all of these models to come together in one unified model to make accurate predictions about Fermi's Paradox, we need an simulation engine to combine them all in one space. We decided to use the Unity engine \cite{Unity} as it offers a graphical environment, where we can visually see exactly what is going on on each planet and how each civilization is expanding.
\smallskip

The model in Unity includes an initial setup as well as 3 different phases:

\begin{enumerate}\addtocounter{enumi}{0}
    \item {\it Initial Setup}

        The Initial conditions of the Galaxy are:
        \begin{enumerate}
            \item The Galaxy, a circular region of arbitrary size, begins with a number \emph{N} of habitable star systems (for simplicity only the habitable planets are shown and are limited to one per star system). This number \emph{N} can be chosen at the beginning of the simulation. (For our simulations, \emph{N} is taken to be 200.)
            \item The planets are spaced out using Poisson Sampling, meaning no two planets will be within a certain minimum distance of each other (also chosen before the simulation begins), creating a well spaced distribution of planets that is still random.
            \item Each of these planets is given a certain probability $\phi_i$ of intelligent life appearing per simulation step, note that $\phi_i$ is different for each planet $i$.
        \end{enumerate}

    \item {\it Pre-Civilization Stage}

    In this 
    Stage, each planet will be continuously running an algorithm that will check, at every time-step (with the aforementioned probability \emph{$\phi_i$}), if the planet has spawned intelligent life or not. If intelligent life succeeds, this planet will move on to the Planetary Civilization stage.
    
    \item {\it Planetary Civilization Stage}

    When any planet reaches this stage, it will begin to run the PET model. Initial values for all of the coefficients and the 3 values of \emph{P}, \emph{E}, and \emph{T} are randomized within a certain realistic range, with a set initial value (for example as taken in \eqref{E:ID}), which is then modified by up to $\pm 10\%$ in either the positive or negative direction.

    Planets can reach the Civilization stage by either evolving intelligent life independently or by being colonized by another intelligent civilization.
    In the latter case, there are two deviations from what was described earlier:

    \begin{enumerate}
    
    \item Initial population \emph{P} (from the PET model's differential equation \eqref{population}) will be set to the number of ``colonists" colonizing the planet.

    \item Technology $T$ will be set to the current technology level of the colonizing civilization, and will be affected not only by the planet in question but also by all planets in the civilization, as described in the next section on the Colonization Stage.

    \end{enumerate}
    
    \item {\it Colonization Stage}

Every civilization originally starts out existing only on one planet. Here, the civilization will have a Population, Environment, and Technology values as described by the PET model.
Once a civilization leaves its home planet, though, to colonize (as defined by the process outlined below), they will have different Population and Environment values for each planet under their control, but the Technology value will be shared amongst every planet in the civilization, such that each planet can contribute to it and every planet's PET model is affected by it.

\end{enumerate}

\smallskip

We next describe the {\it Process of Colonization}. 

Suppose that a civilization 
appeared on planet \emph{A}. 
    
Once this civilization with home planet \emph{A} reaches a threshold of technology and population, they will send out an ``expedition" to colonize another planet \emph{B} (this is selected as the nearest planet to the planet, from which the expedition is sent).
\smallskip

After the expedition is sent, there are two scenarios for what happens to it:
\begin{enumerate}
        \item The expedition will reach planet \emph{B}, which is habitable yet devoid of intelligent life. The colonists from planet \emph{A} will then establish a colony here and begin consuming the planet \emph{B}'s resources, as there is no competition for them.

        \item The expedition reaches planet \emph{B}, but \emph{B} already has an existing intelligent civilization. This will cause direct competition between the 2 civilizations present on this planet, and will eventually lead to conflict.

        This conflict can be simulated using the aforementioned Lanchester Battle Model, with the \emph{fighting effectiveness} of each side determined by that civilization's technology level, and the \emph{strength} of each side determined as such:
        
        \begin{enumerate}
            \item \emph{A}'s strength is the amount of settlers coming from \emph{A},
            
            \item \emph{B}'s strength is a certain percentage of planet \emph{B}'s population (one can think of this percentage as the fraction of \emph{B}'s inhabitants that are in the armed forces).
        
        \end{enumerate}

Once the Lanchester Battle model is run, the winning side will take control of the planet, and will progress with colonization just like in part (a) above.
\end{enumerate}

Planets will only be able to send out a finite number of colony ships, seeing as it is wasteful to attempt to colonize planets around oneself that have already been colonized. This number is similar to the \emph{basic reproductive number} $R_0$ for infections, or the number of uninfected individuals each infected individual spreads the virus to - here instead of spreading viruses between patients, colonies are spread themselves between planets \cite{Delamater2019}. We call this the \emph{$R_0$} value of the civilization, similar to the \emph{$R_0$} value of a virus, and explore the influence of its value on the expansion of colonizations in the next section
    


\section{Results}\label{S:results}

\subsection{One Civilization}
Here, we investigate what changing the \emph{$R_0$} value does to a civilization's growth and how that civilization expands throughout the galaxy. 

For each planet (when it is colonized or a new species emerges on it), we use the same starting conditions as described in Section 2.1.2, however, we vary them slightly to create a stochastic simulation. To do this, we use the process from the {\it Planetary Civilization Stage}~ described in Section 4. This allows us to simulate a range of planets and civilization types.

\begin{figure}[htb]
\begin{subfigure}{.49\textwidth}
\includegraphics[width=1\linewidth,height=0.8\linewidth]{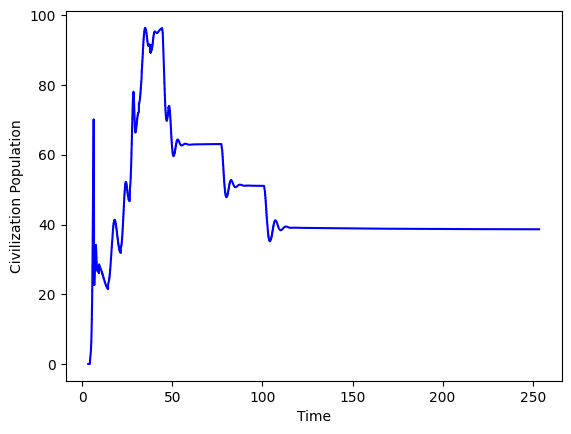}
\subcaption[]{{\footnotesize $R_0 = 2$: Civilization expands briefly then declines. }}
  \label{fig:sub-first}
\end{subfigure}
\begin{subfigure}{.49\textwidth}
\includegraphics[width=1\linewidth,height=0.8\linewidth]{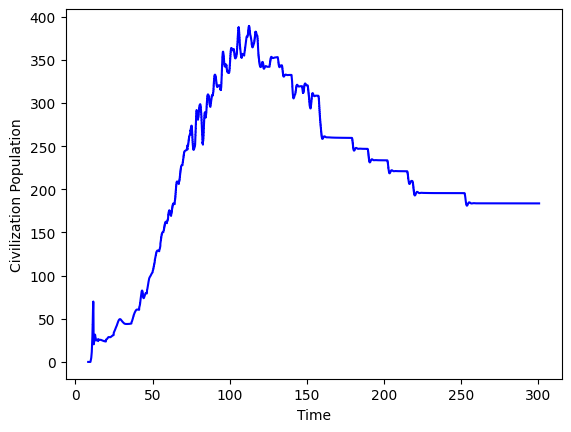}
\subcaption[]{{\footnotesize $R_0 = 3$: Civilization expands widely then declines. }}
  \label{fig:sub-second}
\end{subfigure} \\
\begin{subfigure}{.49\textwidth}
  \includegraphics[width=1\linewidth,height=0.8\linewidth]{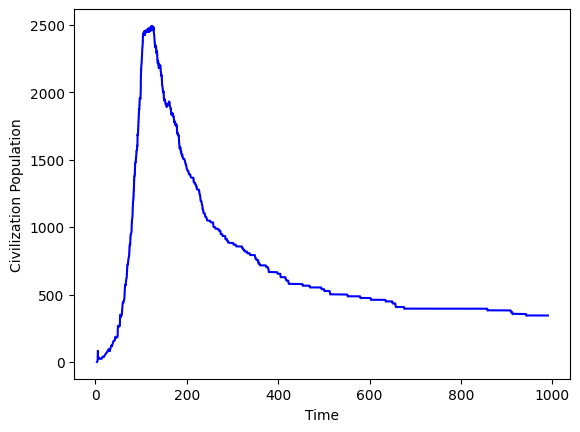}
\subcaption[]{{\footnotesize $R_0 = 4$: Civilization spans galaxy then declines. }}
  \label{fig:sub-third}
\end{subfigure}
\begin{subfigure}{.49\textwidth}
\includegraphics[width=1\linewidth,height=0.8\linewidth]{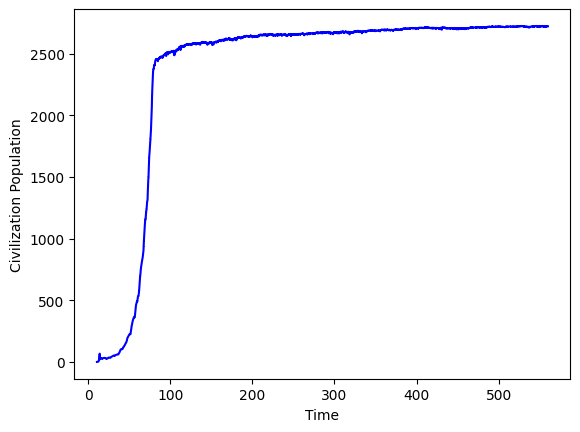}
\subcaption[]{{\footnotesize $R_0 = \infty$: Civilization spans galaxy and persists. }}
  \label{fig:sub-fourth}
\end{subfigure}
\caption{Civilizations with given \emph{$R_0$} values and their colonization success.}
\label{fig:single-civ}
\end{figure}

The graphs in Figure \ref{fig:single-civ} depict the total population of a single civilization over time when left alone in a galaxy with 200 habitable planets. As the civilization colonizes more of the galaxy, its population will grow to fill the planets which it occupies. This total population is obtained by summing the $P$ values from the PET model running on each planet for each time step.

We observe that in the examples of $R_0 = 2,3,4$, the civilizations first expand, but then decline and stabilize, see plots (A), (B), (C) in Figure \ref{fig:single-civ}; the larger the $R_0$ is, the bigger the size of the population after it declines and levels out. Only when $R_0 \to \infty$, the civilization expansion reaches nearly its maximum quite sharply (see plot (D) in Figure \ref{fig:single-civ}) and then does not decline but keeps asymptotically approaching its maximal value. It seems that for a finite $R_0$ any civilization will rise and then decline, no matter what. This makes plausible to investigate the Zoo hypothesis, mentioned in the Introduction \S \ref{S:Zoo}.

\begin{figure}[htb]
\begin{subfigure}{.3\textwidth}
\includegraphics[width=1\linewidth,height=0.85\linewidth]{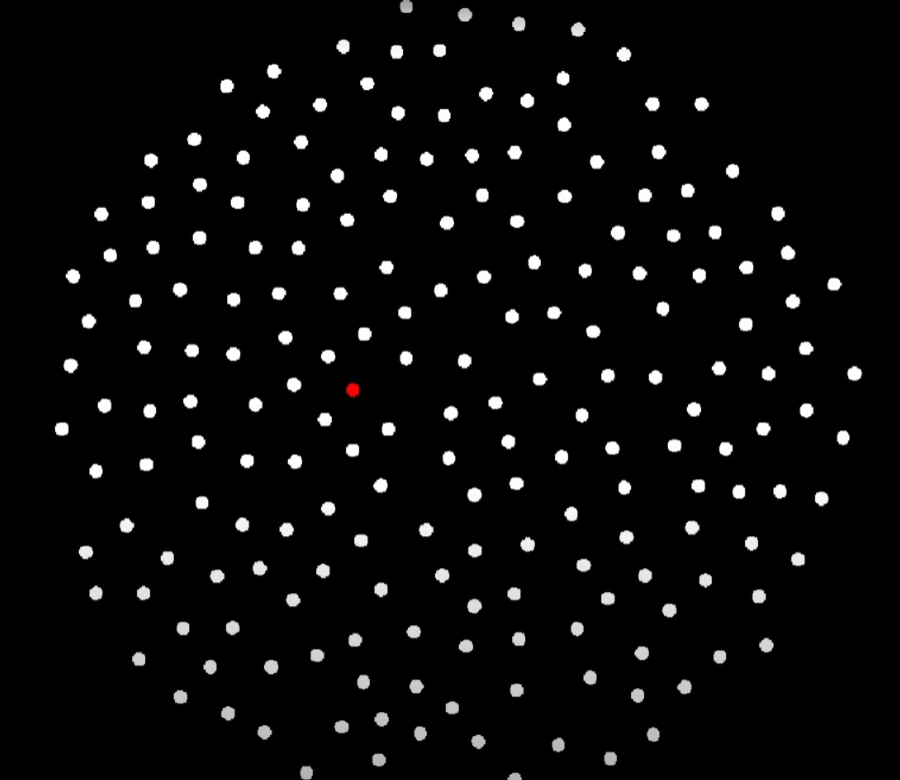}
\subcaption[]{{\footnotesize $t=0$}}
\end{subfigure}
\begin{subfigure}{.3\textwidth}
\includegraphics[width=1\linewidth,height=0.85\linewidth]{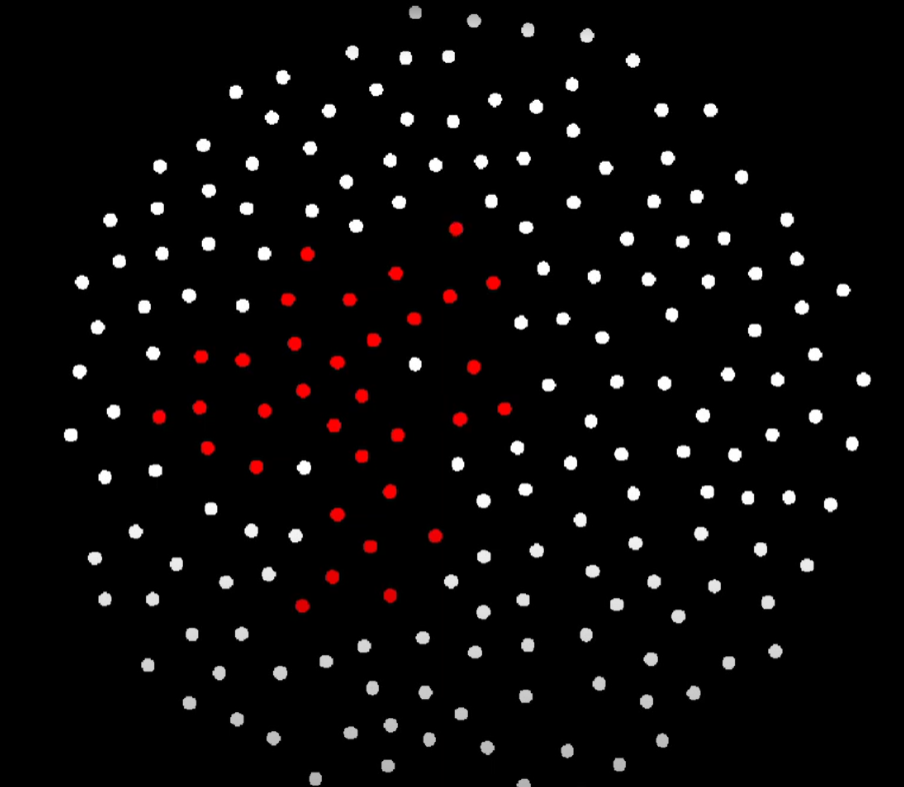}
\subcaption[]{{\footnotesize $t=5$}}
\end{subfigure} 
\begin{subfigure}{.3\textwidth}
  \includegraphics[width=1\linewidth,height=0.85\linewidth]{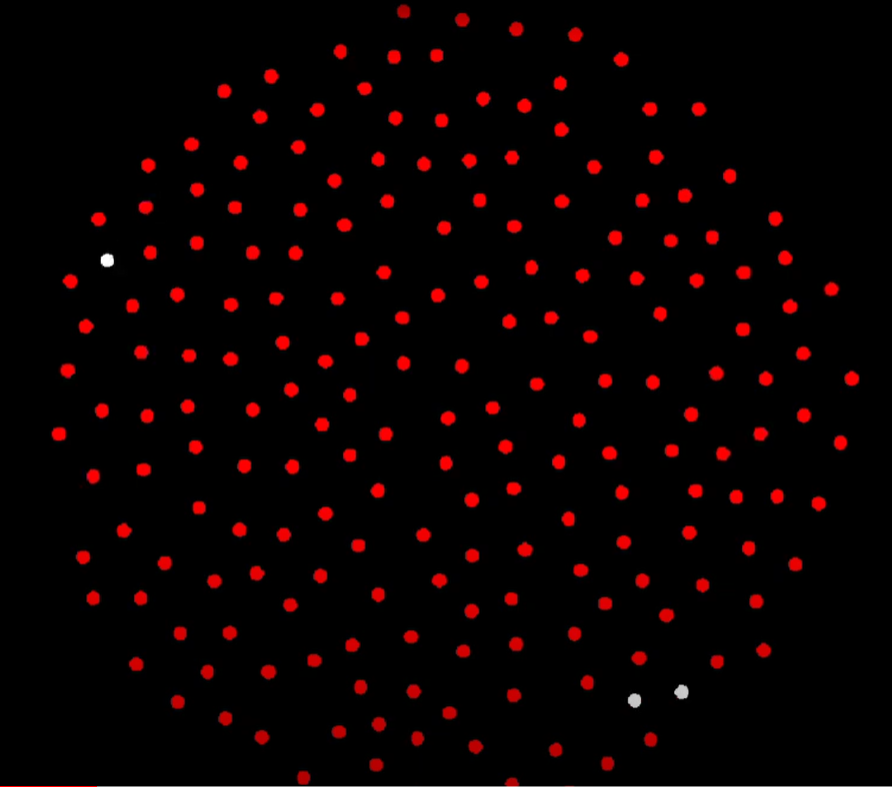}
\subcaption[]{{\footnotesize $t=11$}}
\end{subfigure}\\
\begin{subfigure}{.3\textwidth}
\includegraphics[width=1\linewidth,height=0.85\linewidth]{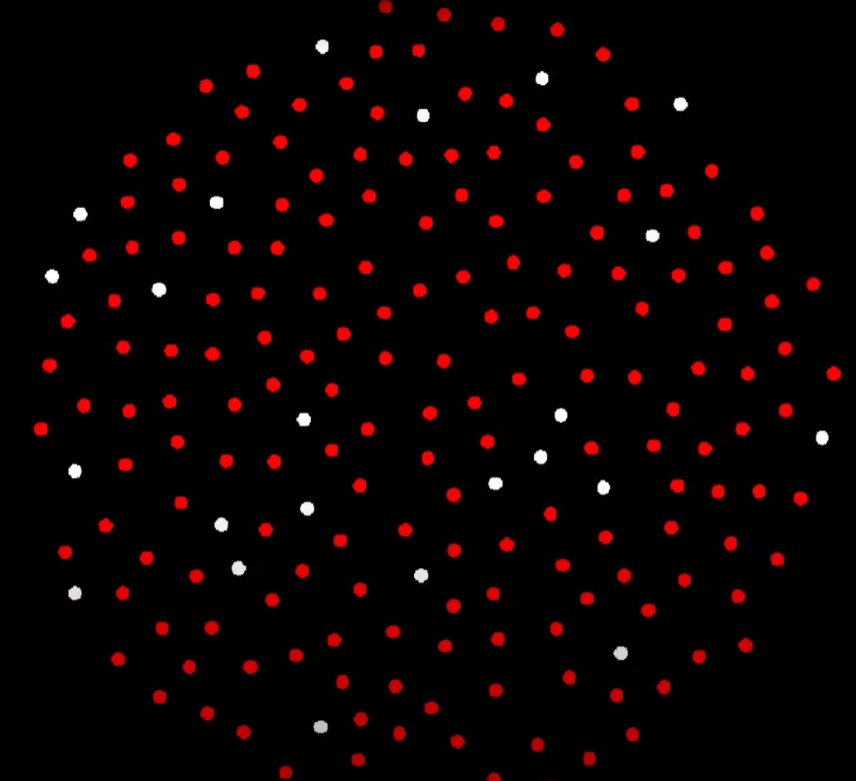}
\subcaption[]{{\footnotesize $t=14$}}
\end{subfigure}
\begin{subfigure}{.3\textwidth}
  \includegraphics[width=1\linewidth,height=0.85\linewidth]{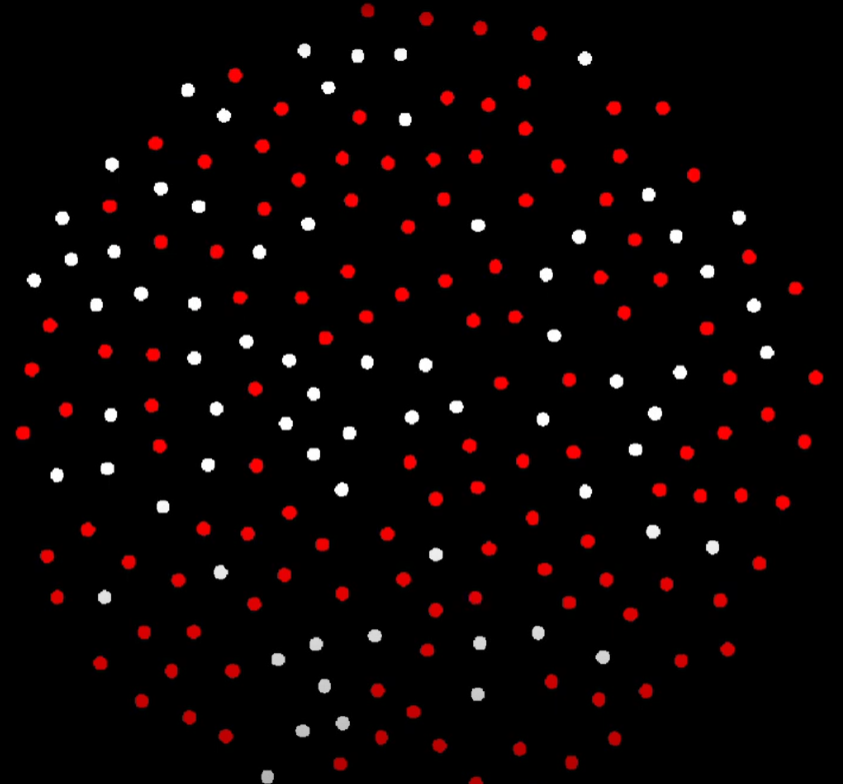}
\subcaption[]{{\footnotesize $t=23$}}
\end{subfigure}
\begin{subfigure}{.3\textwidth}
\includegraphics[width=1\linewidth,height=0.85\linewidth]{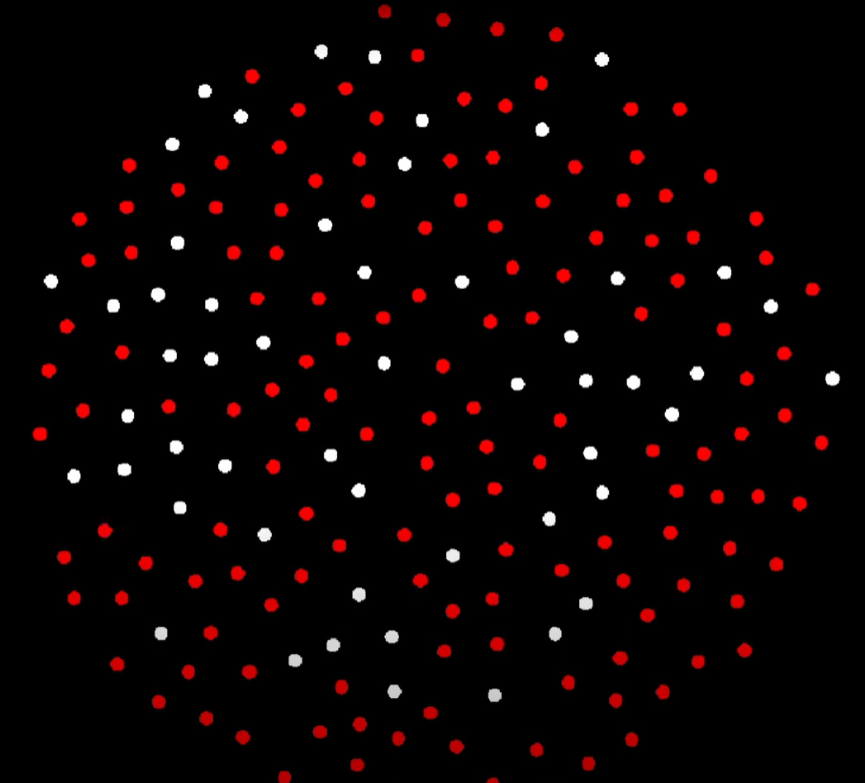}
\subcaption[]{{\footnotesize $t=38$}}
\end{subfigure}
\caption{Evolution of one civilization (with $R_0=4$) that first occupies almost the entire galaxy but then declines to a certain level, as shown in Figure \ref{fig:single-civ}(C), and leaves spatial pockets, possibly  for smaller civilizations to co-exist.}
\label{F:Video}
\end{figure}

To show a visualization of a simulation that shows a possible evolution  for a fixed $R_0$ as graphed in Figure \ref{fig:single-civ}, 
where a civilization first colonizes almost an entire galaxy, as can be seen in plot (C) of Figure \ref{fig:single-civ}, and then declines and stabilizes at a certain level (which could provide a co-existence of other small civilizations), we provide several time snapshots in Figure \ref{F:Video} of the colonization video (available on YouTube, see the link \cite{Video}).  

\newpage

\subsection{Two Civilizations}

Multiple simulations were run with 2 civilizations. Each simulation ended up resulting in one of the following outcomes:
\begin{figure}[htb]
\hspace{-3cm}\includegraphics[width=0.65\textwidth,height=0.4\linewidth]{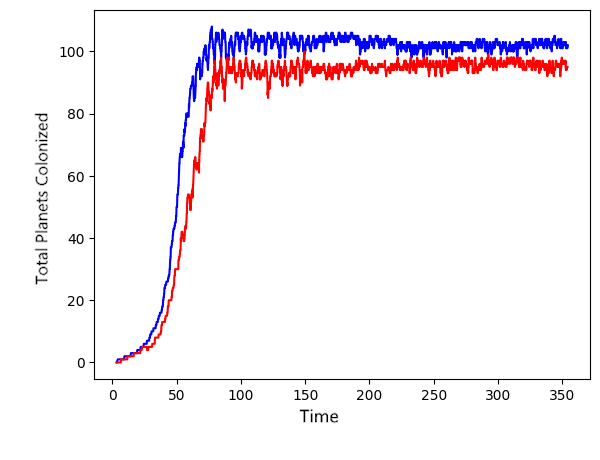}
\caption{Two civilizations start at a similar time, and end up coexisting, each taking roughly half the galaxy.}
\label{F:equal}
\end{figure}

\begin{figure}[htb]
\hspace{-3cm}\includegraphics[width=0.65\textwidth,height=0.4\linewidth]{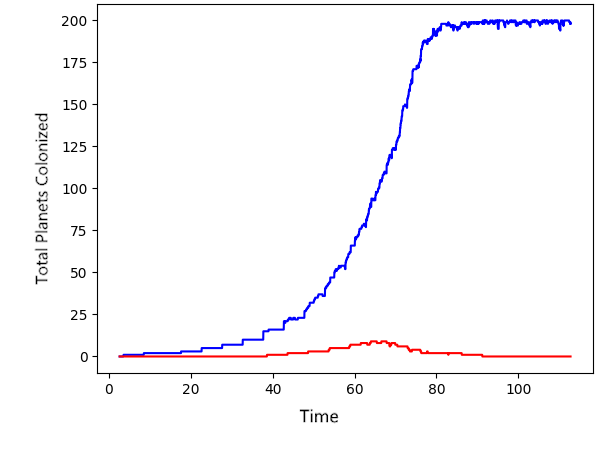}
\caption{Two civilizations emerge at slightly different times, and the older civilization completely wipes out the younger one.}
\label{F:oldwin}
\end{figure}

\begin{figure}[htb]
\hspace{-3cm}\includegraphics[width=0.65\textwidth,height=0.4\linewidth]{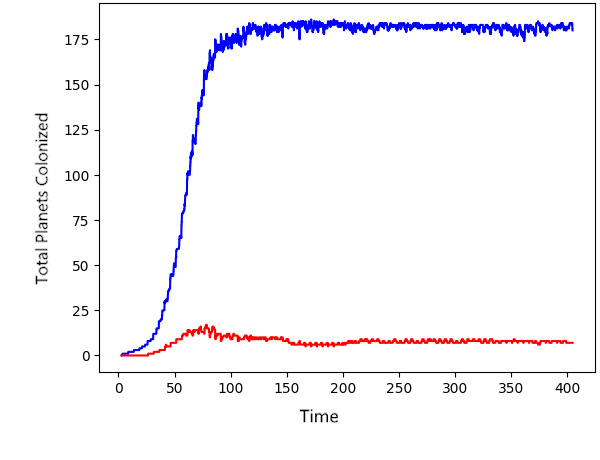}
\caption{Two civilizations emerge at slightly different times, and the older civilization leaves the younger one to exist with a few planets.} 
\label{F:zoo}
\end{figure}

\begin{itemize}
\item
Two civilizations start at a similar time and end up coexisting, each taking roughly half the galaxy, as shown in Figure \ref{F:equal}. 

\item
Two civilizations emerge at slightly different times, and the older civilization completely wipes out the younger one, as shown in Figure \ref{F:oldwin}. 

\item
Two civilizations emerge at slightly different times, and the older civilization leaves the younger one to exist with a few planets, as shown in Figure \ref{F:zoo} (a possible example of the Zoo hypothesis).
\end{itemize}

\smallskip

In the graphs of Figures \ref{F:equal}-\ref{F:zoo}, the $y$-axis depicts number of planets colonized by each civilization (one red and one blue) and the $x$-axis depicts time steps from the start of a simulation.

\newpage

\section{Conclusions}\label{S:Concl}

An unexpected result of our simulations can be seen in Figure \ref{F:zoo}, where {\it a small civilization was left alive by the dominating civilization in the galaxy}. This was due to the large civilization approaching an asymptotic 
state, where each of its populated planets ended up either with no colonization attempts left, no planets near them to colonize, or with planets near them that simply had unsuitable conditions and kept dying out. 

This supports the aforementioned Zoo Hypothesis \cite{wikizoo}, where large civilizations leave smaller ones undisturbed due to a lack of interest in taking them over (a possible evolution scenario of a civilization becoming large and then leaving some planets uncolonized is depicted in Figure \ref{F:Video}).

In our simulation we used an analog of the basic reproductive number from epidemiology \emph{$R_0$}, or the amount of other planets each inhabited planet will send out additional colonies to before ceasing its attempts to colonize. We noticed that this number needed to exceed a critical value of {\it 3 colonized planets per originating planet} for the entire galaxy to be colonized. Otherwise, civilizations would either die out or stagnate, only conquering a small portion of the galaxy and leaving most of it empty. This can be interpreted as civilizations choosing not to send out further expeditions to colonize due to the costs associated with expansion being too high and/or other factors playing into lack of motivation to do so.

We, however, notice that even civilizations with an \emph{$R_0$} of 3 or larger would often leave large pockets of the galaxy un-colonized - after colonizing the galaxy fully, only the planets with the most suitable conditions for life would \emph{remain} colonized. Only when the \emph{$R_0$} was infinite, could the entire galaxy remain under the control of just one civilization indefinitely.

{\textit{\textbf {From these results, we can draw one main conclusion about our own universe:}}

If intelligent extraterrestrial life exists, it may have left empty regions in the galaxy (such as ours) where new civilizations can arise. Intelligent species may already know about our existence but are not interested in us - leaving us alone to develop by ourselves. This also gives insight into our own actions as we expand to a space-faring civilization: perhaps it is best for us to leave other potential civilizations alone instead of invading them like the Europeans did when they colonized the Americas.


\bibliographystyle{acm} 
\bibliography{references}

\end{document}